\documentclass[aps,pra,reprint,superscriptaddress,amsmath,amssymb,floatfix,nofootinbib]{revtex4-2}

\usepackage[colorlinks]{hyperref}
\usepackage{xcolor}
\hypersetup{
    allcolors  = {teal!70!black},
}

\usepackage{graphicx}
\usepackage{bm}
\usepackage{booktabs}
\usepackage{mathtools}
\usepackage{amsthm}
\usepackage{xfrac}

\newcommand{\tr}{\operatorname{tr}}
\newcommand{\R}{\mathbb{R}}
\newcommand{\C}{\mathbb{C}}

\newcommand{\ulm}{Institute of Theoretical Physics, Ulm University, Albert-Einstein-Allee 11, 89081 Ulm, Germany} 
\newcommand{\iqst}{Center for Integrated Quantum Science and Technology (IQST), 89081 Ulm, Germany} 

\newcommand{\inl}{INL -- International Iberian Nanotechnology Laboratory, Av. Mestre Jos\'{e} Veiga s/n, 4715-330 Braga, Portugal}

\newcommand{\uff}{Instituto de F\'{i}sica, Universidade Federal Fluminense, Av. Gal. Milton Tavares de Souza s/n, Niter\'{o}i -- RJ, 24210-340, Brazil}

\usepackage{soul}

\begin{document}

\title{Commutativity from a single Bargmann invariant equality}

\author{Rafael Wagner}
\email{rafael.wagner@uni-ulm.de}
\affiliation{\ulm}
\affiliation{\iqst}

\author{Ernesto F.\ Galv\~ao}
\affiliation{\inl}
\affiliation{\uff}

\begin{abstract}
Noncommutativity of states and observables is a fundamental signature of quantum theory, and a minimal requirement for nonclassicality. We provide a universal necessary and sufficient condition for pairwise commutativity of quantum states $\rho_1$ and $\rho_2$: they commute if and only if $\mathrm{tr}(\rho_1^2\rho_2^2) = \mathrm{tr}(\rho_1 \rho_2 \rho_1 \rho_2)$. For qubits the identity simplifies to an equality between polynomials of purities and of the two-state overlap $\mathrm{tr}(\rho_1\rho_2)$. These multivariate traces---known as Bargmann invariants---are directly measurable, allowing commutativity tests that bypass full state tomography. We point out possible applications to the analysis of POVM simulability and partial photonic distinguishability.
\end{abstract}

\maketitle

\textit{Introduction.---}A fundamental structural question in quantum theory is whether a set of observables acting on a Hilbert space is \emph{jointly diagonalizable}, i.e., whether there exists an orthonormal basis in which they are all diagonal (equivalently, whether they pairwise commute). This problem is relevant in several contexts, including uncertainty relations~\cite{heisenberg1925uber,kennard1927zur,robertson1929uncertainty,maccone2014stronger,catani2022what}, quantumness in multiparameter estimation~\cite{carollo2018uhlmann,carollo2019quantumness,miyazaki2022imaginarityfree}, and positivity criteria for Kirkwood--Dirac quasiprobability representations~\cite{arvidssonshukur2024properties,yunger2018quasiprobability,lostaglio2022kirkwood,gherardini2024quasiprobabilities,wagner2024quantumcircuits,schmid2024kirkwood,wagner2025structuretheoremcomplexvaluedquasiprobability,deBievre2025kirkwood,langrenez2024characterizing,debievre2021complete,deBievre2023relating}. Operationally, noncommutativity of sharp observables has a standard interpretation in terms of joint measurability~\cite{heinosaari2008notesonmeasurability,heinosaari2016invitation,andrejic2020joint,kunjwal2014quantum}: commuting sharp measurements can be jointly implemented.
Sequential measurements thus work as an operational way to establish commutativity, as they will be non-disturbing if and only if they commute.

Whenever the observables are positive, they may also be interpreted as (possibly unnormalized) quantum states. In this case, rather than being operationally associated with measurements, they correspond to preparations (states), whose noncommutativity provides a basis-independent manifestation of quantum coherence~\cite{streltsov2017colloquium,baumgratz2014quantifying,designolle2021set} termed \emph{set coherence}~\cite{designolle2021set}. This notion constitutes the basic primitive underlying quantum resources~\cite{chitambar2019quantum} for collections of states, which have by now been studied in connection with many quantum resources such as entanglement~\cite{zhang2024local,foulds2021controlledSWAP,cai2021entanglement,yu2021absolutely}, coherence~\cite{galvao2020quantum,designolle2021set}, contextuality~\cite{zhang2025quantifierswitnessesnonclassicalitymeasurements,zhang2026reassessingboundaryclassicalnonclassical,rossi2025typicalcontextuality,wagner2024coherence}, Hilbert-space dimension~\cite{bernal2024absolute,galvao2020quantum,giordani2021witnesses,wagner2024inequalities,giordani2023experimental,haakansson2025experimental,santosjunior2026coherence}, imaginarity~\cite{fernandes2024unitary,miyazaki2022imaginarityfree}, non-Gaussianity~\cite{xu2025bargmanninvariantsgaussianstates}, and nonstabilizerness~\cite{wagner2024certifying,zamora2025semi}. In this setting, however, unlike for measurements, no comparably simple necessary and sufficient operational criterion is known for arbitrary collections of preparations, in arbitrary dimensions, in the absence of prior structural information about the states (e.g. purity).

In this Letter, we show that a pair of fourth-order Bargmann invariants~\cite{bargmann1964note,oszmaniec2024measuring,zhang2026surveybargmanninvariantsgeometric,wagner2025coherenceandcontextuality} completely resolves this question: A set $\{\rho_i\}_{i=1}^n$ of (possibly unnormalized) quantum states pairwise commute if and only if 
\begin{equation}\label{eq:criterion_introduction}
\mathrm{tr}(\rho_l^2 \rho_k^2) = \mathrm{tr}(\rho_l \rho_k \rho_l \rho_k), \quad \forall l \neq k\in \{1,\ldots,n\}.
\end{equation}
The same criterion extends to arbitrary quantum observables (i.e. self-adjoint operators). Therefore, set coherence of $\{\rho_i\}_{i=1}^n$ can be decided by direct estimation of the invariants in Eq.~\eqref{eq:criterion_introduction} for all pairs, that is, a total of ${n\choose{2}} =n(n-1)/2 $ fourth-order invariants.

Crucially, Bargmann invariants can be efficiently measured by simple Hadamard-type quantum circuits known as cycle tests ~\cite{oszmaniec2024measuring, quek2024multivariatetrace,oszmaniec2024measuring,simonov2025estimationmultivariatetracesstates,leifer2004measuring,shin2024rankneedestimatingtrace,faehrmann2025intheshadow}. In photonic platforms, moreover, the measurement can be done interferometrically, in a way that is oblivious to the underlying Hilbert space encoding~\cite{pont2022quantifying,jones2020multiparticle,jones2023distinguishability,novo2026nativelinearopticalprotocolefficient,fujii2003exchange,foulds2024generalising,santosjunior2026coherence}. These direct estimation schemes show that any basis-independent criterion we formulate in terms of Bargmann invariants is experimentally accessible, bypassing the high cost of tomography.

We further show that if we know all states live in a 2-dimensional space (qubits), the criterion from Eq.~\eqref{eq:criterion_introduction} reduces to a comparison between second-order Bargmann invariants. Another piece of useful prior information is whether at least one state $\rho_{i_\star}$ in the set $\{\rho_i\}_{i=1}^n$ has no repeated eigenvalues. In this case, the minimal number of invariants required to decide whether there is set coherence is $2(n-1)$, fewer than the $n(n-1)/2$ fourth-order invariants required by Eq.~\eqref{eq:criterion_introduction} whenever $n \geq 3$. Finally, we point out possible applications in simulability of positive operator-valued measures (POVMs), and in the characterization of partial photonic distinguishability. Taken together, these results provide an economical and experimentally meaningful characterization of pairwise commutativity of finite sets of quantum states.

\textit{Background.---}Before proving our main result, we briefly review the existing literature on witnessing noncommutativity of quantum states using Bargmann invariants~\cite{galvao2020quantum,wagner2024inequalities,wagner2026bargmannscenarios,pratapsi2025elementarycharacterizationbargmanninvariants,li2025bargmann,li2025multistateimaginaritycoherencequbit,zhang2024boundaries,xu2025numericalrangesbargmanninvariants,giordani2021witnesses,wagner2025coherenceandcontextuality}, and explain why the resulting criteria are generally only necessary, not sufficient. In this context, the distinction between dimension-dependent criteria and dimension-independent ones is crucial: it highlights the nontriviality of finding dimension-independent necessary-and-sufficient conditions, situating our main result within this literature.

For a fixed set of (possibly unnormalized) quantum states $\{\rho_i\}_{i=1}^n$ on $\mathcal H=\mathbb{C}^d$, let $\vec w=(l_1,\ldots,l_m)$ denote an arbitrary finite sequence of length $m$ with entries in $\{1,\ldots,n\}$. Following Ref.~\cite{designolle2021set}, we say that $\{\rho_i\}_{i=1}^n$ is set incoherent if all pairs $\rho_i$ and $\rho_j$ commute (and set coherent otherwise). We regard $\{1,\ldots,n\}$ as an alphabet of labels, whose elements $l$ are the \emph{letters} and whose finite sequences $\vec w$ are, naturally, \emph{words}. We call any finite set $\mathcal W$ of words a \emph{Bargmann scenario}~\cite{wagner2026bargmannscenarios}. Any word $\vec w$ induces an ordered tuple of states $\rho_{\vec w}\equiv (\rho_{l_1},\ldots,\rho_{l_m})$, which in turn induces the corresponding Bargmann invariant 
\begin{equation}\label{eq:bargmann-def}
\Delta(\rho_{\vec w})
  = \mathrm{tr}\bigl(\rho_{l_1}\rho_{l_2}\cdots\rho_{l_m}\bigr).
\end{equation}
When the tuple of states is either arbitrary or clear from the context, we simply write
\begin{equation}\label{eq:notation}
    \Delta(\rho_{\vec w}) \equiv \Delta_{\vec w} = \Delta_{l_1 \, \cdots \, l_m}.
\end{equation}

For every word $\vec w \in \mathcal{W}$ in a Bargmann scenario, the Bargmann invariant of a tuple of incoherent normalized quantum states $\{\rho_i\}_{i=1}^n$  reduces to~\cite{oszmaniec2024measuring,wagner2024quantumcircuits,wagner2026bargmannscenarios},
\begin{equation}\label{eq:diagonal-bi}
  \Delta_{l_1\cdots l_m}
  = \sum_{\lambda \in \Lambda} p_{\lambda}^{(l_1)}\cdots \,p_{\lambda}^{(l_m)},
\end{equation}
where $p_{\lambda}^{(l_i)} = \langle \lambda \vert \rho_{l_i} \vert \lambda \rangle $ for $\{\vert \lambda \rangle \langle \lambda \vert\}_{\lambda \in \Lambda}$, the orthonormal basis relative to which all states in the set are diagonal. The set of all points $(\Delta_{\vec w})_{\vec w \in \mathcal W}$ achieved by Eq.~\eqref{eq:diagonal-bi} for some set of states in some Hilbert space forms a convex polytope, called the \emph{Bargmann polytope}~\cite{oszmaniec2024measuring,wagner2026bargmannscenarios}, denoted $\mathfrak C(\mathcal{W})$. 

It is immediate that, for any given Bargmann scenario $\mathcal{W}$, membership in $\mathfrak C(\mathcal{W})$ is a \emph{necessary} requirement for joint diagonalizability. However, it is not \emph{sufficient}. To understand this fact, the key ingredient here is the notion of \emph{quantum realizability}~\cite{fraser2023estimationtheoreticapproachquantum,wagner2025coherenceandcontextuality}. We say that an arbitrary $\vec z \in \mathbb{C}^{\mathcal W}$ is quantum realizable in $\mathcal H$ if there exists some set of (possibly unnormalized) states $\{\rho_i\}_{i=1}^n$ realizing it via
\begin{equation}
    z_{\vec w} = \Delta (\rho_{\vec w}),\quad \forall \vec w \in \mathcal W.
\end{equation}
For a given Bargmann scenario $\mathcal{W}$, we denote by $\mathfrak Q(\mathcal{W})$ the set of all points realizable as above, and by $\mathfrak B(\mathcal{W})$ all those realized by \emph{normalized} quantum states. In this case, $\mathfrak C(\mathcal{W})$ corresponds to the points realized by sets of jointly diagonalizable normalized states in some Hilbert space, and it is simple to see that these three sets of correlations satisfy
\begin{equation}
\mathfrak C(\mathcal{W}) \subseteq \mathfrak B(\mathcal{W}) \subseteq \mathfrak Q(\mathcal{W}), \quad \forall \,\mathcal{W}.
\end{equation}

We are now ready to see why membership in $\mathfrak C(\mathcal{W})$ provides necessary, but not sufficient, conditions for set coherence in general. For arbitrary scenarios $\mathcal{W}$, the same point may be quantum realizable both by set incoherent and set coherent collections of states in some Hilbert space. To illustrate this, consider first the Bargmann scenario $\mathcal{W}_3 = \{(1,2,3)\}$. In this case, the associated Bargmann polytope is entirely characterized by $0 \leq z_{123} \leq 1$. What this is saying is that for every $z_{123} \in \mathbb{C}$ satisfying that $z_{123} = z_{123}^*$ and $0 \leq z_{123} \leq 1$, there exists some $\vec \rho  = (\rho_1,\rho_2,\rho_3)$ for which $\{\rho_i\}_{i=1}^3$ are jointly diagonalizable and $z_{123} = \mathrm{tr}(\rho_1 \rho_2\rho_3)$. However, the same point may also be realizable by $\vec \rho$ for which $\{\rho_i\}_{i=1}^3$ is set coherent. Concretely, take the point $z_{123} = 0 \in \mathfrak C(\mathcal{W}_3).$ It can be realized both by $\vec{\sigma} = (\vert 0 \rangle \langle 0 \vert,\vert 2 \rangle \langle 2 \vert, \vert 4 \rangle \langle 4 \vert)$ via $$\Delta(\sigma_{123}) = \langle 0 \vert 2 \rangle \langle 2 \vert 4 \rangle \langle 4 \vert 0 \rangle = 0$$ or by $\vec{\sigma}' = (\vert 0 \rangle \langle 0 \vert, \vert + \rangle \langle + \vert , \vert 2 \rangle \langle 2 \vert )$ via $$\Delta(\sigma_{123}') = \langle 0 \vert +\rangle \langle + \vert 2 \rangle \langle 2 \vert 0 \rangle  = 0,$$ 
where we have denoted $\vert + \rangle = \sfrac{1}{\sqrt{2}}(\vert 0 \rangle + \vert 1 \rangle)$. The first is a set-incoherent realization in $\mathbb{C}^5$, while the second is a set-coherent realization in $\mathbb{C}^3$.

To give another example, we can consider the Bargmann scenario $\mathcal{W}_{C_3} = \{(1,2),(1,3),(2,3)\}$ which is an instance of the graph-theoretic formalism from Refs.~\cite{galvao2020quantum,wagner2024inequalities}. In this case, one can show that the Bargmann polytope $\mathfrak C(\mathcal{W}_{C_3})$---which in this case is referred to as the \emph{event-graph polytope}~\cite{wagner2024inequalities,wagner2025coherenceandcontextuality}---is given by $z_{ij} = z_{ij}^*$, $0\leq z_{ij} \leq 1$ for all $i<j$, together with the inequalities
\begin{subequations}
\begin{align}
    +z_{12}+z_{13}-z_{23}\leq 1,\\
    +z_{12}-z_{13}+z_{23}\leq 1,\\
    -z_{12}+z_{13}+z_{23}\leq 1.
\end{align}
\end{subequations}
These inequalities are the  facets of this Bargmann polytope~\cite{ziegler1995lectures,brondsted1983introduction}. As before,  $\vec z = (z_{12},z_{13},z_{23}) \in \mathfrak C(\mathcal{W}_{C_3})$ implies that there exists some list $\vec \rho = (\rho_1,\rho_2,\rho_3)$ of jointly diagonalizable states for which $z_{ij} = \Delta(\rho_{ij})=\mathrm{tr}(\rho_i\rho_j)$. Again, one can easily find points that can also be realized by set coherent states, for example the point $\vec z = (\sfrac{1}{2},\sfrac{1}{2},\sfrac{1}{2})$. 

It is immediate that for most Bargmann scenarios membership in $\mathfrak C(\mathcal{W})$ does not guarantee pairwise commutativity for every quantum realization. This is somewhat expected given that these constraints are \textit{system-independent}, that is, valid for every possible Hilbert space realizing the points via some specific tuple of states. However, once one fixes additional structure---such as the Hilbert-space dimension or whether the states are pure---there exist known necessary and sufficient criteria formulated solely in terms of Bargmann invariants. For a trivial instance, if the states are known to be pure, then two states are set coherent if and only if they are not both elements of the same orthonormal basis; equivalently, for two pure states $\psi_1,\psi_2$ one has that $\Delta_{12} \notin \{0,1\}$.

Let us now consider Bargmann invariants of quantum states of a specific dimension. For qubits, Ref.~\cite{li2025multistateimaginaritycoherencequbit} provided a Gram matrix criterion which also admits a simple geometric formulation. A set of normalized states $\{\rho_i\}_{i=1}^n$ is set incoherent iff the corresponding Bloch vectors are all collinear. Writing a single-qubit state as $\rho = (\mathbb{I}+\langle \vec{\mathrm r}_\rho , \vec{{P}}\rangle )/2,$ where $\vec{P} = (X,Y,Z)$ denotes the vector of Pauli matrices, and defining for $\vec \rho = (\rho_1,\ldots,\rho_n)$ the Gram matrix 
\begin{equation}\label{eq:Gram_of_blochs}
    G_{\vec{\mathrm{r}}_{\vec \rho}} = (\langle \vec{\mathrm{r}}_{\rho_i},\vec{\mathrm{r}}_{\rho_j}\rangle )_{ij}
\end{equation}
one finds that $\{\rho_i\}_{i=1}^n$ is set incoherent iff $\mathrm{rank}(G_{\vec{\mathrm{r}}_{\vec \rho}}) = 1$. In higher dimensions, an analogous construction can be given in terms of generalized Bloch vectors built from the generalized Gell--Mann basis~\cite{kimura2003bloch,bertlmann2008bloch}, but it yields only a necessary condition: $\mathrm{rank}(G_{\vec{\mathrm{r}}_{\vec \rho}}) \leq d-1$~\cite{li2025multistateimaginaritycoherencequbit}.

None of the constructions we have reviewed obtain necessary and sufficient, dimension-independent conditions for set coherence. 
Refs.~\cite{oszmaniec2024measuring,wagner2026bargmannscenarios} pointed out that for every \emph{fixed} $n$ and $d$ there are, in principle, Bargmann scenarios $\mathcal{W}_{n,d}$ for which membership in $\mathfrak C(\mathcal{W})$ decides whether the sets of states realizing the points are coherent or not, i.e. $\vec z \in \mathfrak C(\mathcal{W})$ iff all states $\vec \rho$ realizing $\vec z = \vec \Delta(\vec \rho\,)$ are jointly diagonalizable. Their construction of scenarios is dimension-dependent and fairly suboptimal (as the minimal number of words in $\mathcal{W}$ depends on the Hilbert space dimension $d$). 

\textit{Universal criterion.---}We now show, however, that there is one special Bargmann scenario, namely 
\begin{equation}
\mathcal{W}_{\rm inc}^{(2)} = \{(1,1,2,2),(1,2,1,2)\},
\end{equation}
that completely characterizes set coherence for pairs of quantum states $\{\rho_1,\rho_2\}$ in arbitrary finite-dimensional Hilbert spaces $\mathcal{H}$ (and, more broadly, commutativity for arbitrary sets of quantum observables via its straightforward extensions). In such scenarios, membership in the Bargmann polytope defined by 
\begin{equation}\label{eq:Bargmann_scenario_equality}
\mathfrak C(\mathcal{W}_{\rm inc}^{(2)}) = \left\{(z_{1122},z_{1212}) \in [0,1]^2 \mid z_{1122}=z_{1212}\right\}
\end{equation}
is necessary and sufficient for set incoherence of any quantum realization, i.e. $\rho_1$ and $\rho_2$ are set incoherent iff $\mathrm{tr}(\rho_1^2\rho_2^2)=\mathrm{tr}(\rho_1\rho_2\rho_1\rho_2)$. 

The proof is almost immediate and holds for arbitrary self-adjoint operators acting on a Hilbert space, so we state it in this more general form. Let $A_1$ and $A_2$ be any pair of observables acting on a finite-dimensional Hilbert space $\mathcal{H}$. Recall that for arbitrary operators $F,G$ such that $F-G$ is self-adjoint, one has~\cite{quek2024multivariatetrace}
$$
\Vert F - G\Vert_2^2 := \mathrm{tr}(\vert F-G\vert^2)
= \mathrm{tr}(F^2)+\mathrm{tr}(G^2)-2\mathrm{tr}(FG).
$$
Setting $F = iA_1A_2$ and $G = iA_2A_1$, so that $F-G=i[A_1,A_2]$, and using that $\Vert X \Vert_2^2 = 0$ if and only if $X=0$, we find that
\begin{equation}
    [A_1,A_2] = 0 \quad \text{iff}\quad \tr(A_1^2A_2^2)-\tr(A_1A_2A_1A_2) = 0.
\end{equation}
For the case where $A_1$ and $A_2$ are both positive operators---so that we may interpret them as (possibly unnormalized) quantum states---we conclude that
\begin{equation}\label{eq:main_equation}
[\rho_1,\rho_2] = 0 \quad \text{iff} \quad  \Delta_{1122}=\Delta_{1212};
\end{equation}
Note, moreover, that the relation with the $\Vert \cdot \Vert_2$ norm implies that $\mathrm{tr}(\rho_1^2\rho_2^2)\geq\mathrm{tr}(\rho_1\rho_2\rho_1\rho_2)$ for all states.
 
For normalized states, Eq.~\eqref{eq:main_equation} is precisely the Bargmann scenario from Eq.~\eqref{eq:Bargmann_scenario_equality}, thus showing that, for arbitrary pairs of quantum states $\vec \rho = (\rho_1,\rho_2)$ in arbitrary finite-dimensional Hilbert spaces,   $(\Delta(\rho_{1122}),\Delta(\rho_{1212})) \in \mathfrak C(\mathcal{W}_{\rm inc}^{(2)})$ if and only if $\{\rho_1,\rho_2\}$ commute. For possibly unnormalized states, it shows furthermore that $$\mathrm{Cone}[\mathfrak C(\mathcal{W}_{\rm inc}^{(2)})] = \{\lambda \vec z \mid \lambda \geq 0, \vec z \in \mathfrak C(\mathcal{W}_{\rm inc}^{(2)})\}$$ yields the set incoherent realizations within the convex cone of unnormalized states $\mathfrak Q(\mathcal{W}_{\rm inc}^{(2)})$.

Our criterion can also be used to bound particular manifestations of set coherence, such as \emph{set imaginarity}~\cite{miyazaki2022imaginarityfree}. A set of states $\{\rho_i\}_{i=1}^n$ is said to be real-representable (or imaginarity-free) if there exists an orthonormal basis in which all density matrices have only real entries. Otherwise, the set is said to possess imaginarity. For any sequence $\vec \rho = (\rho_l,\rho_k,\rho_s)$ drawn from $\{\rho_i\}_{i=1}^n$, the imaginary part of the third-order Bargmann invariant $\Delta_{lks}$, a witness of set imaginarity~\cite{fernandes2024unitary}, satisfies
\begin{equation}
    2 \vert \mathrm{Im}[\Delta_{l k s}] \vert \leq \sqrt{\Delta_{ll}}\sqrt{2(\Delta_{k k ss}-\Delta_{k s k s})},
\end{equation}
which follows from the Cauchy--Schwarz inequality in the form $\vert \mathrm{tr}(F^\dagger G) \vert \leq \Vert F \Vert_2 \Vert G \Vert_2$ applied to $$2i\mathrm{Im}[\Delta_{l k s}]=\mathrm{tr}(\rho_l[\rho_k,\rho_s]),$$ and from $\Vert [\rho_k,\rho_s]\Vert_2^2=2(\Delta_{k k ss}-\Delta_{k s k s})$.

In Table~\ref{tab:criteria} we organize the conditions we have reviewed in comparison with the new criterion of Eq.~\eqref{eq:main_equation}. In
End Matter~\ref{app:incomparability} we show that the first three conditions are incomparable, i.e., that one can find examples of set coherent triplets detected by violation of only one condition, thus showing they witness different manifestations of how a set of states can be coherent.

\begin{table}[t]
\centering
\footnotesize
\setlength{\tabcolsep}{4pt}
\begin{tabular}{c@{\hskip 4pt}l@{\hskip 4pt}cc}
 \hline
 \hline
 & Condition on Bargmann invariants & Nec. & Suff. \\
 \hline
 & $\Delta_{\vec w}\in [0,1]$, $\forall \vec w$~\cite{oszmaniec2024measuring,fernandes2024unitary}
  & Yes & No \\
& Event-graph polytope facets~\cite{galvao2020quantum,wagner2024inequalities}
  & Yes & No \\
& $\mathrm{rank}(G_{\vec{\mathrm{r}}_{\vec \rho}})\leq d{-}1$~\cite{li2025multistateimaginaritycoherencequbit} & Yes & $d\!=\!2$ \\
& $\Delta_{l l k k}=\Delta_{l k l k}\ \forall\,l \neq k $
    (Eq.~\eqref{eq:main_equation}) & Yes & Yes \\
 \hline
 \hline
\end{tabular}
\caption{Necessary or sufficient conditions for joint
diagonalizability of $n$ normalized states. The only system-dependent condition, appearing in the third row, assumes a $2$-dimensional Hilbert space.}
\label{tab:criteria}
\end{table}

While Eq.~\eqref{eq:main_equation} constitutes the central result of this Letter, the underlying identity relating multivariate traces to the commutator norm has already appeared previously in different contexts. As already noted, Ref.~\cite{quek2024multivariatetrace} highlights this basic connection between the norm $\Vert \cdot \Vert_2$ and multivariate traces. A further link arises through the Wigner--Yanase--Dyson skew information~\cite{wigneryanase1963information,girolami2014observable},
$$
I_{\sfrac{1}{2}}(F,G)=
\tr(F G^2)-\tr(F^{\sfrac{1}{2}} G F^{\sfrac{1}{2}} G),
$$
which, upon choosing $F = A_1^2$ (so that $F^{\sfrac{1}{2}} = A_1$) and $G = A_2$, reproduces the same structural expression. Wigner and Yanase originally observed that this quantity vanishes if and only if $[F,G]=0$. More recently, Ref.~\cite{girolami2014observable} interpreted it as a measure of coherence (in its standard basis-dependent formulation) of a quantum state $\rho$ with respect to the eigenbasis of a Hamiltonian $H$. Here we place this quantity in a different context by linking it to nonclassicality via its quantum realizability in terms of Bargmann invariants, thereby extending the associated criterion in a manner that is independent of the Hilbert space dimension.

\textit{Simpler conditions from partial information on the states.---}We now investigate whether prior information on the states realizing the invariants may help in simplifying the criterion from Eq.~\eqref{eq:main_equation}. Ref.~\cite{li2025multistateimaginaritycoherencequbit} showed that, for any set of single-qubit normalized states $\{\rho_i\}_{i=1}^n$, the only nontrivial information provided by higher-order Bargmann invariants is  the sign of their imaginary part. That is, for a given invariant $\Delta_{\vec w} = x_{\vec w} \pm i y_{\vec w}$, the relevant information is whether one has $x_{\vec w} + i y_{\vec w}$ or $x_{\vec w} - i y_{\vec w}$. In particular, real-valued Bargmann invariants such as $\Delta_{1122}$ or $\Delta_{1212}$ can be fully characterized by polynomials in two-state overlaps (see End Matter~\ref{app:criterion_qubits}): 
\begin{align}
    \Delta_{1122}&=\Delta_{12}+\frac{1}{2}(\Delta_{11}\Delta_{22} -1)\\
    \Delta_{1212}&=\Delta_{12}^2+\frac{\Delta_{11}+\Delta_{22}-\Delta_{11}\Delta_{22}-1}{2}
\end{align}
Using this observation, one finds that Eq.~\eqref{eq:main_equation} reduces, for sets of single-qubit states, to the condition
\begin{equation}\label{eq:qubit_nonlinear_criterion_1}
    \left(\Delta_{12}-\frac{1}{2}\right)^2= \left(\Delta_{11}-\frac{1}{2}\right)\left(\Delta_{22}-\frac{1}{2}\right).
\end{equation}
Alternatively, another way to see that this provides a necessary and sufficient condition for the qubit case is to use the criterion $\mathrm{rank}(G_{\vec{\mathrm{r}}_{\vec \rho}})= 1$ from Eq.~\eqref{eq:Gram_of_blochs}, which in this situation is equivalent to the vanishing of all $2\times 2$ principal minors of the Gram matrix of Bloch vectors, leading to the same equation.

This reformulation reflects a more general phenomenon: once the Hilbert space dimension is fixed, the linear constraints arising from facets of Bargmann polytopes translate into \emph{nonlinear} constraints on lower-order invariants. A similar behavior is known to occur under purity assumptions, in situations where higher-order Bargmann invariants can, in some cases, be reconstructed from complete sets of lower-order ones~\cite{oszmaniec2024measuring,avdoshkin2023extrinsic}. More generally, the constraints derived in Ref.~\cite{li2025multistateimaginaritycoherencequbit} provide a systematic route for converting arbitrary linear facet-defining inequalities bounding the Bargmann polytope $\mathfrak C(\mathcal{W})$ into nonlinear equalities or inequalities involving two-state overlaps, which can serve as nonlinear witnesses of set coherence for the single-qubit case.

It is also possible to show that considering the Bargmann invariants $\Delta_{1122}$ and $\Delta_{1212}$ is \emph{crucial} for testing set coherence of two states $\{\rho_1,\rho_2\}$ of dimension $d \geq 4$. Consider the Bargmann scenario \begin{align*}
\mathcal{W}=\bigr\{(1,1),(1,1,1),(2,2),(2,2,2),\\ (1,2),(1,1,2),(1,2,2)\bigr\}
\end{align*}
having all 2 and 3 letter words from the alphabet $\{1,2\}$. It is possible to show explicitly (see End Matter~\ref{app:irreducible}), that there is a point $\vec z \in \mathfrak C(\mathcal{W})$ realized both by a pair of noncommuting states $(\rho_1,\rho_2)$ and by a pair of commuting states $(\sigma_1,\sigma_2)$ (as required for membership in $\mathfrak C(\mathcal{W})$). 

Since Eq.~\eqref{eq:main_equation} gives a necessary and sufficient criterion for set coherence in arbitrary dimensions, a set of $n$ states $\{\rho_i\}_{i=1}^n$ can be tested by estimating the $\binom{n}{2}$ pairwise fourth-order Bargmann invariants. This can be reduced when prior structural information is available. For instance, if one state $\rho_{i^\star}$ is known to have a fully non-degenerate spectrum, then it suffices to test only the $n-1$ commutators $[\rho_{i^\star},\rho_i]$ with $i\neq i^\star$, requiring $2(n-1)$ fourth-order Bargmann invariants. Indeed, in this case commutativity of all $\rho_i$ with $\rho_{i^\star}$ implies pairwise commutativity of the whole set.

\textit{Discussion and outlook.---}In this Letter, we have shown that noncommutativity of sets of quantum states---a basis-independent formulation of quantum coherence for a set of states---admits a complete operational characterization in terms of a single equality between fourth-order Bargmann invariants: For arbitrary pairs $\{\rho_1,\rho_2\}$ in arbitrary Hilbert spaces, commutativity is equivalent to $\mathrm{Tr}(\rho_1^2\rho_2^2)= \mathrm{Tr}(\rho_1\rho_2\rho_1\rho_2)$. This immediately yields a universal, basis-independent criterion for set coherence as introduced by Designolle \textit{et al.}~\cite{designolle2021set}. Additionally, we also clarify how the required number of invariants can be reduced when additional structure is available, such as prior information on Hilbert space dimension, purity, or whether some reference state has a fully non-degenerate spectrum. More broadly, this simple criterion places set coherence on the same footing as other operationally accessible quantum signatures, since it is expressed entirely in terms of experimentally measurable multivariate traces of states. In particular, direct estimation of the relevant invariants is much less resource-intensive than, for instance, doing full tomography of all states (a point also raised by Ref.~\cite{girolami2014observable}).  

Our construction also applies to sets of unnormalized states and thus yields a necessary and sufficient criterion for the set coherence of POVMs $\{E_k\}_{k=1}^n$~\cite{designolle2021set}. As shown in Ref.~\cite{designolle2021set}, POVMs with no set coherence lie in the convex hull of projectively simulable measurements~\cite{oszmaniec2017simulating}. A natural direction for future work is to use our criteria to quantify set coherence in POVMs and to explore whether this leads to operationally meaningful links between set coherence, projective simulability, and measurement disturbance.

We expect these results to find concrete applications well beyond the foundational characterization developed here. One particularly natural arena is the analysis of partial distinguishability in multiphoton linear interferometers~\cite{shchesnovich2015partial,shchesnovich2018collective,pont2022quantifying,menssen2017distinguishability,rodari2025semi,rodari2024experimentalobservationcounterintuitivefeatures,seron2023boson,giordani2020experimental,giordani2021witnesses,brod2019witnessing}. Recent work has emphasized that distinguishability and mixedness can affect quantum interference in qualitatively different ways, including distinctions between triplets of states that are colinear in the Bloch ball and triplets that span a finite volume~\cite{jones2023distinguishability}. Our criterion gives a complementary, basis-independent handle on this problem: it can distinguish incoherent partial distinguishability, where the internal states are jointly diagonalizable, from genuinely coherent distinguishability, where the states may be mixed yet fail to commute. This aligns closely with the framework of incoherent distinguishability recently developed in Ref.~\cite{annoni2025incoherentbehaviorpartiallydistinguishable}, which isolates the regime in which multiphoton distinguishability behaves as a stochastic error, and how this affects the output statistics of linear interferometers, which are known to depend functionally on Bargmann invariants of the internal quantum states of the incoming single photons~\cite{shchesnovich2015partial,shchesnovich2018collective}.

More generally, we view our criterion as a contribution to the broader programme of using unitary invariants to characterize quantum nonclassicality in a basis-independent way~\cite{wagner2024quantumcircuits,wagner2026bargmannscenarios,wagner2024inequalities,wagner2024certifying,wagner2023anomalous}. Besides the foundational appeal,  the simple measurement schemes for these invariants suggest new feasible experiments for certifying quantum nonclassical resources  \cite{giordani2020experimental,giordani2021witnesses,giordani2023experimental}. 

\textit{Acknowledgments.---}We thank Rui Soares Barbosa, Leonardo Novo, and Dario Cilluffo for helpful discussions. We acknowledge the use of Claude Opus 4.6 (Anthropic), GPT-5.4 Thinking-Mini, and of hybrids of the DeepSeek V3 series as an aid in exploring and verifying segments of all mathematical proofs, polishing, condensing, and editing this manuscript. The final writing, the mathematical results, their derivations, and their correctness were independently verified by the authors, who take full responsibility for the technical accuracy and integrity. R.W.\ acknowledges support from the Alexander von Humboldt Foundation. E. F. G.  acknowledges support from the National Council for Scientific and Technological Development -- CNPq (Brazil) under grant 308292/2025-1.


\appendix

\section*{End Matter}

\subsection{Incomparability of criteria}\label{app:incomparability}

We exhibit tuples of normalized states showing that the three families of necessary conditions listed in Table~\ref{tab:criteria} witness \emph{different} manifestations of set coherence of quantum states. The witnesses listed were:
\begin{enumerate}
    \item\label{item:criteria_i} Reality and non‑negativity of all Bargmann invariants realizable by set incoherent states.
    \item\label{item:criteria_ii} Bounds from facet inequalities from  event‑graph polytopes  (Bargmann polytopes for two-letter words $(l,k)$ such that $l \neq k$).
    \item \label{item:criteria_iii} Given a certain fixed dimension $d$, the Gram matrix of the Bloch vectors $G_{\vec{\mathrm{r}}_{\vec \rho}} = (\langle \vec{\mathrm{r}}_{\rho_i},\vec{\mathrm{r}}_{\rho_j}\rangle )_{ij}$ has rank smaller than or equal to $d-1$. 
\end{enumerate}
We will do so by showing each criterion can witness a form of set coherence that the other criteria cannot. As an important remark, note that these three criteria \emph{are not} on equal footing. The first two criteria work regardless of prior information of the state in terms of its Hilbert space dimension, while the last one requires this dimension to be known. 

\textit{Set coherence witnessed only by ~\ref{item:criteria_i}.---} 
Take the three pure single-qubit states,
$$\rho_1 = \lvert 0\rangle\langle 0 \vert ,\quad \rho_2=\lvert+\rangle\langle + \vert ,\quad  \rho_3 = \lvert+_i\rangle\langle +_i \vert ,$$ whose set coherence is witnessed by
$\Delta(\rho_{123})=(1+i)/4\notin\R_{\geq 0}$~\cite{fernandes2024unitary,oszmaniec2024measuring}. As mentioned in Ref.~\cite{wagner2024quantumcircuits} all pairwise
overlaps equal $1/2$ which is a point inside every event-graph polytope $\mathfrak C(\mathcal{W}_G)$, for arbitrary event graphs $G$ as it is the convex combination of $\vec 0$ and $\vec 1$ which are also in $\mathfrak C(\mathcal{W}_G)$. Moreover, embedding these states in $\mathbb{C}^4$, the associated Gram matrix of Bloch vectors has
eigenvalues $(5/4,1/2,1/2)$, so its rank is exactly $d-1=3$ and not larger.  

\textit{Set coherence witnessed only by~\ref{item:criteria_ii}.---}That~\ref{item:criteria_i} and~\ref{item:criteria_ii} are not comparable was already pointed out in Ref.~\cite{wagner2024quantumcircuits}. We now recall this point and extend it to the third criterion as well. Three pure single-qubit states whose Bloch vectors lie on
a common great circle of the Bloch sphere given by 
$$\vec{\mathrm{r}}_{\rho_1}=(1,0,0),\,\, 
\vec{\mathrm{r}}_{\rho_2}=(\tfrac12,\tfrac{\sqrt 3}{2},0),\,\, \vec{\mathrm{r}}_{\rho_3}=(\tfrac12,-\tfrac{\sqrt 3}{2},0),$$ forming the celebrated \emph{trine ensemble}~\cite{barnett2009quantumstatediscrimination,galvao2020quantum,chaturvedi2026epistemic,halder2025quantumadvantage,halder2024identifying}, have all
third-order Bargmann invariants real and non-negative. Nevertheless, they maximally violate the
event-graph inequality
via the relation $z_{12}+z_{13}-z_{23}
= \mathrm{tr}(\rho_1\rho_2)+\mathrm{tr}(\rho_1\rho_3)-\mathrm{tr}(\rho_2\rho_3) = \tfrac34+\tfrac34-\tfrac14=\tfrac54>1$ discussed in the main text~\cite{galvao2020quantum}. Proceeding as previously, if we embed the trine in $\C^4$ the same
example also satisfies~\ref{item:criteria_iii}, since in this case the rank of the Gram matrix of Bloch vectors satisfies $\mathrm{rank}(G_{\vec{\mathrm{r}}_{\vec \rho}})\leq 3\leq
d-1$. In other words, this is an example of a set coherent tuple whose noncommutativity is witnessed by~\ref{item:criteria_ii} and not by~\ref{item:criteria_i} or~\ref{item:criteria_iii}.

\textit{Set coherence witnessed only by~\ref{item:criteria_iii}.---}Let us now fix the Hilbert space dimension to be $\mathbb C^4$, and consider the four rank-2 mixed states
\begin{align*}
  \rho_k &= \tfrac12\bigl(|1\rangle\langle 1|
           +|k{+}1\rangle\langle k{+}1|\bigr),\quad k=1,2,3,\\
  \rho_4 &= \tfrac12\bigl(|a\rangle\langle a|+|b\rangle\langle b|\bigr),
\end{align*}
with $|a\rangle=(|1\rangle+|3\rangle)/\sqrt 2$ and
$|b\rangle=(|2\rangle+|4\rangle)/\sqrt 2$. A direct computation gives
$\mathrm{tr}(\rho_{i}^2)=\sfrac{1}{2}$ and $\mathrm{tr}({\rho_i\rho_j})= \sfrac{1}{4}$ for every $i\neq j \in \{1,\ldots,4\}$. This implies that these states cannot violate facet-defining inequalities of  event graphs since, as before, they are convex combinations of $\vec 0$ and $\vec 1$ which are in $\mathfrak C(\mathcal{W}_G)$ for any event graph $G$. Moreover, it is also simple to show that every
higher-order Bargmann invariant is non-negative and real (for instance, the first ones are  
$\mathrm{tr}({\rho_1\rho_2\rho_3})= \sfrac{1}{8}$, $\mathrm{tr}({\rho_1\rho_2\rho_4})=\mathrm{tr}({\rho_1 \rho_3 \rho_4}) = \mathrm{tr}({\rho_2 \rho_3 \rho_4})=\sfrac{1}{16}$,
$\mathrm{tr}({\rho_1 \rho_2 \rho_3 \rho_4})= \sfrac{1}{32}$). In turn, the Gram matrix of generalized Bloch vectors equals $G_{\mathrm{\vec{r}}_{\vec \rho}} = \sfrac{1}{4}\,\mathbb{I}_4$ which has rank 4. 

\subsection{Criterion for pairs of qubits}\label{app:criterion_qubits}

For a single-qubit state $\rho$ with Bloch vector $\vec{\mathrm{r}}_\rho\in\mathbb{R}^3$ ($\Vert \vec{\mathrm{r}}_\rho \Vert \le 1$), we write
\begin{equation}
\rho = \frac{1}{2}\bigl(\mathbb I_2 + \langle \vec{\mathrm{r}}_\rho,\vec{P}\rangle \bigr),
\end{equation}
so that $\operatorname{tr}(\rho_i\rho_j) = \frac12\bigl(1+\langle \vec{\mathrm{r}}_{\rho_i}, \vec{\mathrm{r}}_{\rho_j}\rangle \bigr)$.  
In the Appendix of Ref.~\cite{li2025multistateimaginaritycoherencequbit}, one finds the following expressions for the real and imaginary parts of arbitrary Bargmann invariants $\Delta_{1234} = \operatorname{tr}(\rho_1\rho_2\rho_3\rho_4)$ for single qubit states:
\begin{equation}
\operatorname{tr}(\rho_1\rho_2\rho_3\rho_4) = \frac{a_0^{(4)}+i\,b_0^{(4)}}{2^{3}},
\end{equation}
with
\begin{align}
a_0^{(4)} =& +(1+\langle \vec{\mathrm{r}}_{\rho_1}, \vec{\mathrm{r}}_{\rho_2}\rangle )(1+\langle \vec{\mathrm{r}}_{\rho_3}, \vec{\mathrm{r}}_{\rho_4}\rangle ) \nonumber 
           \\&- (1-\langle \vec{\mathrm{r}}_{\rho_1},\vec{\mathrm{r}}_{\rho_3}\rangle )(1-\langle \vec{\mathrm{r}}_{\rho_2},\vec{\mathrm{r}}_{\rho_4}\rangle ) \nonumber 
           \\&+ (1+\langle \vec{\mathrm{r}}_{\rho_1},\vec{\mathrm{r}}_{\rho_4}\rangle )(1+\langle \vec{\mathrm{r}}_{\rho_2},\vec{\mathrm{r}}_{\rho_3}\rangle ), \\
b_0^{(4)} =& \det\bigl(\vec{\mathrm{r}}_{\rho_1}+\vec{\mathrm{r}}_{\rho_2},\; \vec{\mathrm{r}}_{\rho_2}+\vec{\mathrm{r}}_{\rho_3},\; \vec{\mathrm{r}}_{\rho_3}+\vec{\mathrm{r}}_{\rho_4}\bigr). 
\end{align}
Since in our case the relevant Bargmann invariants are real, for both ordered quadruples $(\rho_1,\rho_1,\rho_2,\rho_2)$ and $(\rho_1,\rho_2,\rho_1,\rho_2)$ we have $b_0^{(4)} = 0$. Let us calculate first $\Delta_{1122}$. In this case, $\vec{\mathrm{r}}_{\rho_1}=\vec{\mathrm{r}}_{\rho_2} \equiv \vec{\mathrm{r}}_{A}$ and $\vec{\mathrm{r}}_{\rho_3}=\vec{\mathrm{r}}_{\rho_4}\equiv \vec{\mathrm{r}}_{B}$ 
and we introduce, furthermore, the symbols $
A = \Vert \vec{\mathrm{r}}_A \Vert ^2,\,\, B = \Vert \vec{\mathrm{r}}_B \Vert ^2$, and $ C = \langle \vec{\mathrm{r}}_A,\vec{\mathrm{r}}_B\rangle . $
Inserting these into the equation for $a_0^{(4)}$ we get 
\begin{align*}
a_0^{(4)} &= (1+A)(1+B) - (1-C)^2 + (1+C)^2\\
&=1+A+B+AB + 4C.
\end{align*}
Expressing $A,\,\,B,$ and $C$ in terms of two-state overlaps out of the relation between overlaps and the inner-product between Bloch vectors via $A = 2\Delta_{11}-1$, $B = 2\Delta_{22}-1$ and $C = 2\Delta_{12}-1$ we find, after a series of  simplifications,
\begin{align}
\operatorname{tr}(\rho_1^2\rho_2^2)&=\frac{a_0^{(4)}}{8}= \frac{1+A+B+AB+4C}{8} \nonumber \\
&=\Delta_{12}+\frac{1}{2}(\Delta_{11}\Delta_{22}-1).
\end{align}

Proceeding similarly for the ordered quadruple $(\rho_1,\rho_2,\rho_1,\rho_2)$ we find 
\begin{align*}
a_0^{(4)} &= (1+C)^2 - (1-A)(1-B) + (1+C)^2\\
&= 1+A+B+4C+2C^2-AB,
\end{align*}
and therefore
\begin{align}
\operatorname{tr}(\rho_1\rho_2\rho_1\rho_2)&=\frac{a_0^{(4)}}{8}
= \frac{1+A+B+4C+2C^2-AB}{8}\nonumber \\
&=\Delta_{12}^2+\frac{1}{2}(\Delta_{11}+\Delta_{22}-\Delta_{11}\Delta_{22}-1).
\end{align}

\subsection{Irreducibility for higher dimensions}\label{app:irreducible}

Here, we construct a pair of noncommuting states $\rho_1$ and $\rho_2$ in $\mathcal{H}=\mathbb C^4$ sharing all lower order Bargmann invariants with another pair of commuting states $\sigma_1$ and $\sigma_2$, thus showing that at least for $d\geq 4$ one cannot conclude set coherence without considering higher-order invariants. Let us consider 
\begin{align}
    \rho_1 &= \mathrm{diag}(\sfrac{1}{2},\sfrac{3}{8},\sfrac{1}{8},0),\\
    \rho_2 &= \mathrm{diag}(\sfrac{4}{15},\sfrac{1}{3},\sfrac{1}{6},\sfrac{7}{30})+\sfrac{1}{10}\,R, 
\end{align}
where $R = \vert 1 \rangle \langle 3 \vert + \vert 3 \rangle \langle 1 \vert + \vert 2 \rangle \langle 4 \vert + \vert 4 \rangle \langle 2 \vert$ is a traceless Hermitian operator. By construction the two operators do not commute, captured by our criterion since 
\begin{equation}
    \mathrm{tr}(\rho_1^2\rho_2^2)-\mathrm{tr}(\rho_1\rho_2\rho_1\rho_2)=\sfrac{9}{3200}>0.
\end{equation}
Let us now consider a different pair of states given by 
\begin{align}
    \sigma_1 &= \mathrm{diag}(\sfrac{1}{2},\sfrac{3}{8},\sfrac{1}{8},0),\\
    \sigma_2 &= \mathrm{diag}(\sfrac{11}{30},\sfrac{2}{15},\sfrac{11}{30},\sfrac{2}{15}).
\end{align}
One can then check that for these examples
\begin{equation}
\begin{gathered}
  \mathrm{tr}(\rho_1^2)=\mathrm{tr}(\sigma_1^2)=\tfrac{13}{32},\quad
  \mathrm{tr}(\rho_1^3)=\mathrm{tr}(\sigma_1^3)=\tfrac{23}{128},\\
  \mathrm{tr}(\rho_2^2)=\mathrm{tr}(\sigma_2^2)=\tfrac{137}{450},\quad
  \mathrm{tr}(\rho_2^3)=\mathrm{tr}(\sigma_2^3)=\tfrac{31}{300},\\
  \mathrm{tr}(\rho_1\rho_2)=\mathrm{tr}(\sigma_1\sigma_2)=\tfrac{67}{240},\quad
  \mathrm{tr}(\rho_1^2\rho_2)=\mathrm{tr}(\sigma_1^2\sigma_2)=\tfrac{223}{1920},\\
  \mathrm{tr}(\rho_1\rho_2^2)=\mathrm{tr}(\sigma_1\sigma_2^2)=\tfrac{653}{7200},
\end{gathered}
\end{equation}
which implies that the set coherence of $\{\rho_1,\rho_2\}$ cannot be witnessed by this set of Bargmann invariants (i.e., it would realize a point inside the Bargmann polytope of the associated list of 2 and 3 letter words).

\bibliography{references}

\end{document}